\documentclass[11pt]{article}
\usepackage[dvips]{graphicx}
\author{K. Van Acoleyen\footnote{karel.vanacoleyen@rug.ac.be} and H. Verschelde}
\title{Dynamical mass generation by source inversion: Calculating the mass gap of the Gross-Neveu model.}
\def\beb{\begin{equation}}
\def\bee{\end{equation}}
\def\bea{\begin{eqnarray}}
\def\eea{\end{eqnarray}}
\def\gg#1{g^{#1}}
\def\l#1{\overline{#1}}
\def\b#1{\beta_{#1}}
\def\g#1{\gamma_{#1}}
\def\du{\mu\frac{\partial}{\partial\mu}}

\def\bb#1{b_{#1}}
\def\f#1#2{\frac{#1}{#2}}
\def\X#1{X_{#1}}
\def\Y#1{Y_{#1}}
\def\A#1{A_{#1}}
\def\B#1{B_{#1}}
\def\gb{\f{\g{0}}{2\b{0}}}
\begin{document}
\maketitle
\begin{center}Department of Mathematical Physics and Astronomy,\\
University of Ghent,\\
 Krijgslaan 281,9000 Ghent,Belgium
\end{center}
\section*{Abstract}
We probe the U(N) Gross-Neveu model with a source-term
$J\l{\Psi}\Psi$. We find an expression for the renormalization
scheme and scale invariant source $\widehat{J}$, as a function of
the generated mass gap. The expansion of this function is
organized  in such a way that all scheme and scale dependence is
reduced to one single parameter d. We get a non-perturbative mass
gap as the solution of $\widehat{J}=0$. In one loop we find that
any physical choice for d gives good results for high values of N.
In two loops we can determine d self-consistently by the principle
of minimal sensitivity and find remarkably accurate results for
$N>2$.
\newpage
\section{Introduction}
Asymptotic free massless field theories suffer from
I.R.-renormalons, which signal that one is expanding around a
wrong vacuum. To cure the theory from these renormalons the
particles must acquire a mass. The only dimensionful parameter
available is $\Lambda_{\overline{MS}}$. A problem arises: in
conventional perturbation theory $\Lambda_{\overline{MS}}$ appears
only as a log not as a power.

In the recent past, two new solutions, besides the
$1/N$-expansion, have been formulated for this problem in the
context of the Gross-Neveu model \cite{gn}, which is a very
appealing model to test non-perturbative methods because the exact
result for the mass gap \cite{forgacs} is known. One of the two
solutions consisted of a renormalizable version of the Optimized
Expansion\cite{arvanitz}, while the other, found by one of us,
presented a fully renormalizable action for the local composite
operator $\l{\Psi}\Psi$ \cite{henri},\cite{henrii}. Both methods
require some tricks to maintain renormalizability.

In this paper we present a third method which has the advantage of
a straightforward renormalization. The idea is very simple: we add
a source term $J\overline{\Psi}\Psi$ to the Gross-Neveu
Lagrangian, then we calculate the mass gap to obtain the relation
$m(J)$. After inversion\footnote{The idea of inversion is not new,
it was already formulated by Fukuda\cite{fuk1}. He considered the
inversion of $\Phi(J)$, where $\Phi$ is a composite operator. An
extensive review of the method and its applications can be found
in \cite{fuk2}.}, we find an expression of $\widehat{J}$, the
renormalization scheme and scale independent quantity associated
with $J$, as a function of $m$. Putting $\widehat{J}(m)$ equal to
zero, gives us all possible solutions for the mass gap.
Furthermore we will organize the expansion of $\widehat{J}(m)$ in
such a way, that all the scheme and scale dependence reduces to
one single parameter $d$. This parameter is fixed by the principle
of minimal sensitivity.

In section \ref{sec:meth}, the method is explained in detail.
Section \ref{sec:numerical}, contains the numerical results of the
one and two-loop calculations.

\section{The Method}\label{sec:meth}
Let us perturb the $U(N)$ Gross-Neveu model in $2-\epsilon$
dimensions with a $\overline{\Psi}\Psi$ composite operator. \beb
\mathcal{L}=\overline{\Psi}i\partial^{\mu}\gamma_{\mu}\Psi+\frac{1}{2}\mu^{\epsilon}g^{2}{(\overline{\Psi}\Psi)}^{2}-J\overline{\Psi}\Psi+\mathcal{L}_{\textrm{c.t.}}
\label{lagr}\bee We have $N$ Dirac fermions which enjoy a manifest
$U(N)$-symmetry. This theory is asymptotically free and for $J=0$,
possesses a discrete $\g{5}$ invariance which in perturbation
theory leads to a vanishing mass gap. It is also renormalizable
with the counterterms: \beb
\mathcal{L}_{\textrm{c.t.}}=i\delta_{Z}\overline{\Psi}\partial^{\mu}\gamma_{\mu}\Psi+\frac{1}{2}\mu^{\epsilon}\delta_{{Z}_g}g^{2}{(\overline{\Psi}\Psi)}^{2}
-\delta_{{Z}_2}J\overline{\Psi}\Psi. \bee These counterterms
depend on the scale $\mu$ and on the renormalization scheme.
Because we are interested in the limit $J\rightarrow0$, we will
restrict ourselves to the mass independent renormalization
schemes. In the context of the mass gap, every (mass independent)
R.S., can be obtained from another (the $\overline{MS}$-scheme for
instance) by the following transformations: \begin{eqnarray}
\overline{J}=J(1+a_{0}g^{2}+a_{1}g^{4}+\ldots)\label{jtr},\\
\overline{g}^{2}=\gg{2}(1+b_{0}\gg{2}+b_{1}\gg{4}+\ldots)\label{gtr},
\end{eqnarray} $a_{i},b_{i}$ are finite numbers and
$\overline{J}=\overline{J}(\mu),J=J(\mu),\overline{g}^{2}=\overline{g}^{2}(\mu),\gg{2}=\gg{2}(\mu)$
is understood. From now on we take the convention that
$\l{J},\l{g}^{2},\l{\b{i}},\l{\g{i}}$,\ldots, refer to the
$\l{MS}$ scheme, while $J,\gg{2},\b{i},\g{i}$ refer to an
arbitrary R.S.. The scale dependence of $J$ and $\gg{2}$ is
governed by the $\gamma$ and $\beta$- functions: \begin{eqnarray}
\du
J|_{J_{0},g_{0},\epsilon}\equiv-J\gamma(\gg{2})&=&-J(\g{0}\gg{2}
+\g{1}\gg{4}+\g{2}\gg{6}+\ldots)\label{gamma},\\
\du
\gg{2}|_{g_{0},\epsilon}\equiv\beta(\gg{2})&=&-2(\b{0}\gg{4}+\b{1}\gg{6}+\b{2}\gg{8}+\ldots)\label{beta}.\end{eqnarray}
The R.S.-dependence of $\g{1},\g{2},\b{2}$
($\g{0}=\l{\g{0}},\b{0}=\l{\b{0}},\b{1}=\l{\b{1}}$) can be easily
calculated using (\ref{jtr})-(\ref{beta}). In the formulation of
our method we were led by the requirement that physical results
must be independent of the scale and the scheme.

We now define the scale and scheme independent $\widehat{J}$:\beb
\widehat{J}\equiv J f(\gg{2}),\label{defJ}\bee with \beb \du
f(\gg{2})|_{g_{0},\epsilon}=\gamma(\gg{2}) f(\gg{2})\label{deff}.
\bee The scale independence is seen immediately. As a solution of
(\ref{deff}), we find:\begin{eqnarray}{f(\gg{2})=\lambda
g^{-\f{\g{0}}{\b{0}}}(1-\f{1}{2}(\f{\g{1}}{\b{0}}-\f{\g{0}\b{1}}{\b{0}^{2}})\gg{2}
+(-\f{\g{2}}{4\b{0}}+\f{\g{1}\b{1}}{4\b{0}^{2}}-\f{\g{0}}{4\b{0}}(\f{\b{1}}{\b{0}})^{2}+\f{\g{0}\b{2}}{4\b{0}^2}}\nonumber\\
+\f{\g{1}^{2}}{8\b{0}^{2}}-\f{\g{1}\g{0}\b{1}}{4\b{0}^{3}}+\f{\g{0}^{2}\b{1}^{2}}{8\b{0}^{4}})\gg{4}+\ldots),\label{f}
\end{eqnarray}with $\lambda$ an integration constant. We choose $\lambda=1$, any
other choice implies a different
$\widehat{J}:\widehat{J}(\lambda)=\lambda \widehat{J}(0)$. This is
not a problem because we will look for solutions of
$\widehat{J}=0$. One can now easily check the scheme independence
of $\widehat{J}$.

In order to get $\widehat{J}(m)$, we will first have to calculate
$m(J)$. A perturbative calculation of the mass gap generated by
(\ref{lagr}) will give us:\beb
m(J)=J(1+\gg{2}(\X{0}'+\X{1}'\ln\f{J^{2}}{\mu^{2}})+\gg{4}(\Y{0}'+\Y{1}'\ln\f{J^{2}}{\mu^{2}}+\Y{2}'(\ln\f{J^{2}}{\mu^{2}})^{2})+\ldots).
\bee After inversion, we arrive at a relation of the following
form:\begin{eqnarray}
J(m)=m(1+\gg{2}(\X{0}+\X{1}\ln\f{m^{2}}{\mu^{2}})\nonumber\\
+\gg{4}(\Y{0}+\Y{1}\ln\f{m^{2}}{\mu^{2}}+\Y{2}(\ln\f{m^{2}}{\mu^{2}})^{2})+\ldots).
\end{eqnarray}The coefficients which multiply a log can easily be related
to the other coefficients and the coefficients of the $\gamma$ and
$\beta$ functions. Demanding (\ref{gamma}) and (\ref{beta}), we
obtain:\begin{eqnarray}
J(m)=m(1+\gg{2}(\X{0}+\f{\g{0}}{2}\ln\f{m^{2}}{\mu^{2}})+\gg{4}(\Y{0}+(\f{\g{1}}{2}+(\f{\g{0}}{2}-\b{0})\X{0})\ln\f{m^{2}}{\mu^{2}}\nonumber\\
+\f{1}{4}(\f{\g{0}^{2}}{2}-\b{0}\g{0})(\ln\f{m^{2}}{\mu^{2}})^{2})+\ldots)\label{genj}.\end{eqnarray}
Plugging this result together with (\ref{f}), in (\ref{defJ}) we
finally arrive at:

\begin{eqnarray}
\widehat{J}&=&m(g^{-\f{\g{0}}{\b{0}}})\Bigg[1+\gg{2}\bigg(\X{0}-\f{1}{2}(\f{\g{1}}{\b{0}}-\f{\g{0}\b{1}}{\b{0}^{2}})+\f{\g{0}}{2}\ln \f{m^{2}}{\mu^{2}}\bigg)\nonumber\\
&&+\gg{4}\bigg(\f{\X{0}}{2}(\f{\g{0}\b{1}}{\b{0}^{2}}-\f{\g{1}}{\b{0}})-\f{\g{2}}{4\b{0}}+\f{\g{1}\b{1}}{4\b{0}^{2}}-\f{\g{0}}{4\b{0}}(\f{\b{1}}{\b{0}})^{2}+\f{\g{0}\b{2}}{4\b{0}^2}
+\f{\g{1}^{2}}{8\b{0}^{2}}\nonumber\\
&&-\f{\g{1}\g{0}\b{1}}{4\b{0}^{3}}+\f{\g{0}^{2}\b{1}^{2}}{8\b{0}^{4}}+\Y{0}
+\Big(\f{\g{0}}{4}(\f{\g{0}\b{1}}{\b{0}^{2}}-\f{\g{1}}{\b{0}})+\f{\g{1}}{2}+(\f{\g{0}}{2}-\b{0})\X{0}\Big)\ln\f{m^{2}}{\mu^{2}}\nonumber \\
&&+\f{1}{4}\Big(\f{\g{0}^{2}}{2}-\b{0}\g{0}\Big)(\ln\f{m^{2}}{\mu^{2}})^{2}\bigg)+\ldots \Bigg]\\
& \equiv&
m(g^{-\f{\g{0}}{\b{0}}})\Big[1+\gg{2}(\A{0}+\A{1}\ln\f{m^{2}}{\mu^{2}})+\gg{4}(\B{0}+\B{1}\ln\f{m^{2}}{\mu^{2}}\nonumber\\
&&+\B{2}(\ln\f{m^{2}}{\mu^{2}})^{2})+\ldots \Big]\label{Jtilde}
\end{eqnarray}
Due to the scheme independence of $\widehat{J}$, the coefficients
$\A{i},\B{i}$ are independent of the mass renormalization
(\ref{jtr}). They depend only on the coupling constant
renormalization (\ref{gtr}). We find \footnote{The fastest way to
arrive at (\ref{ABtr}), is the substitution of (\ref{gtr}) in
formula (\ref{Jtilde}), written in the $\overline{MS}$-scheme.
Alternatively one can also use the scheme-dependence of
$\g{1},\g{2},\b{2},\X{0}$ and $\Y{0}$, checking explicitly the
mass-renormalization independence.} \bea \A{0}&=&\l{\A{0}}-\bb{0}
\f{\g{0}}{2\b{0}}\nonumber\\\A{1}&=&\l{\A{1}}\nonumber\\
\B{0}&=&\l{\B{0}}+\f{\g{0}}{4\b{0}}(\f{\g{0}}{2\b{0}}
+1){\bb{0}}^{2}-\f{\g{0}}{2\b{0}}\bb{1}+\bb{0}\l{\A{0}}(1-\f{\g{0}}{2\b{0}})\label{ABtr}\\
\B{1}&=&\l{\B{1}}+\bb{0}\l{\A{1}}(1-\f{\g{0}}{2\b{0}})\nonumber\\
\B{2}&=&\l{\B{2}}.\nonumber\eea So the expansion of $\widehat{J}$
in powers of $\gg{2}$ is still highly scheme
$(\gg{2},\bb{0},\bb{1})$ and scale $(\mu)$ dependent. We now show
that this dependence is reduced to one simple parameter $d$, by
using another expansion. Following an argument of Grunberg
\cite{grunberg}(II,A), we know that $\widehat{J}$, a scheme and
scale-invariant quantity, depends only on $m$ and
$\Lambda_{\l{MS}}$ : $\widehat{J}\equiv
mF\big(\f{m}{\Lambda_{\l{MS}}}\big)$. Reorganizing (\ref{Jtilde})
as an expansion in powers of $\f{1}{\b{0}\ln
\f{m^{2}}{{\Lambda_{\l{MS}}}^{2}}}$ will leave one arbitrariness:
we can equally well expand in $\f{1}{\b{0}\ln
\f{m^{2}}{{\Lambda_{\l{MS}}}^{2}}+d}$. To show this explicitly we
will need some formulas:\bea \gg{2}(\mu)&=&\f{1}{ \b{0}\ln
\f{\mu^{2}}{{\Lambda}^{2}}}\Bigg(1-\f{\b{1}}{\b{0}}\f{\ln \big(\ln
\f{\mu^{2}}{{\Lambda}^{2}}\big)}{\b{0}\ln
\f{\mu^{2}}{{\Lambda}^{2}}}\nonumber\\
&&+\f{(\f{\b{1}}{\b{0}})^{2}\bigg(\Big(\ln\big(\ln
\f{\mu^{2}}{{\Lambda}^{2}}\big)\Big)^{2}-\ln\big(\ln
\f{\mu^{2}}{{\Lambda}^{2}}\big)
\bigg)+\Big(\f{\b{2}}{\b{0}}-\f{{\b{1}}^{2}}{{\b{0}}^{2}}\Big)}{(\b{0}\ln
\f{\mu^{2}}{{\Lambda}^{2}})^{2}}\nonumber\\
&&+\mathcal{O}\Big(\f{1}{\b{0}\ln
\f{\mu^{2}}{{\Lambda}^{2}}}\Big)^{3}\Bigg) \label{rung}\\
\Lambda&=&\Lambda_{\l{MS}}\exp[-\f{\bb{0}}{2\b{0}}]\qquad\textrm{(see
e.g. \cite{celmaster})} \label{ltr}\\
\b{2}&=&(\bb{0}^{2}-\bb{1})\b{0}+\b{1}\bb{0}+\l{\b{2}}\qquad\textrm{(see
\cite{grunberg})}. \eea Using these formulas\footnote{Note that
the expansion (\ref{rung}) is only valid for large positive values
of $\b{0}\ln\f{{\mu}^{2}}{\Lambda^{2}}$. } together with
(\ref{Jtilde}) and (\ref{ABtr}), we find the master formula:\bea
\widehat{J}&=&m(\b{0}\ln
\f{m^{2}}{{\Lambda_{\l{MS}}}^{2}}+d)^{\f{\g{0}}{2\b{0}}}\times\nonumber\\&&\Bigg[1
+\f{1}{(\b{0}\ln
\f{m^{2}}{{\Lambda_{\l{MS}}}^{2}}+d)}\bigg[\l{\A{0}}+\f{\g{0}\b{1}}{2{\b{0}}^{2}}\ln(\ln
\f{m^{2}}{{\Lambda_{\l{MS}}}^{2}}+\f{d}{\b{0}})-\f{d\g{0}}{2\b{0}}\bigg]\nonumber\\
&&+\f{1}{(\b{0}\ln
\f{m^{2}}{{\Lambda_{\l{MS}}}^{2}}+d)^{2}}\bigg[\l{\B{0}}+\l{\A{0}}(\gb-1)\f{\b{1}}{\b{0}}\ln(\ln
\f{m^{2}}{{\Lambda_{\l{MS}}}^{2}}+\f{d}{\b{0}})\nonumber\\
&&+(\f{\b{1}}{\b{0}})^{2}\Big([\ln(\ln
\f{m^{2}}{{\Lambda_{\l{MS}}}^{2}}+\f{d}{\b{0}})]^{2}\gb(\f{\g{0}}{4\b{0}}-\f{1}{2})+\gb\ln(\ln
\f{m^{2}}{{\Lambda_{\l{MS}}}^{2}}+\f{d}{\b{0}})\Big)\nonumber\\
&&-\gb(\f{\l{\b{2}}}{\b{0}}-\f{{\b{1}}^{2}}{{\b{0}}^{2}})+d^{2}\big(\f{\g{0}}{4\b{0}}(\gb-1)\big)\nonumber\\
&&+d\Big(\l{\A{0}}(1-\gb)-\f{\g{0}\b{1}}{2{\b{0}}^{2}}+\f{\g{0}\b{1}}{2{\b{0}}^{2}}\ln(\ln
\f{m^{2}}{{\Lambda_{\l{MS}}}^{2}}+\f{d}{\b{0}})(1-\gb)
\Big)\bigg]\nonumber\\
&&+\mathcal{O}\Big(\f{1}{\b{0}\ln
\f{m^{2}}{{\Lambda_{\l{MS}}}^{2}}+d}\Big)^{3}
\Bigg],\label{Jtildeln} \eea with
$d\equiv\bb{0}-\b{0}\ln\f{m^{2}}{{\mu}^{2}}$.

We can now recover the original Gross-Neveu model, by putting
$J_{0}$, the naked source, equal to zero:\beb
J_{0}(m)\sim\widehat{J}(m)=0\bee One finds a non-perturbative mass
gap, that is a solution of \bea
F(\f{m}{\Lambda_{\l{MS}}})&=&(\b{0}\ln
\f{m^{2}}{{\Lambda_{\l{MS}}}^{2}}+d)^{\gb}\bigg[1+\f{1}{(\b{0}\ln
\f{m^{2}}{{\Lambda_{\l{MS}}}^{2}}+d)}\Big[\ldots\Big]\nonumber\\
&&+\f{1}{(\b{0}\ln
\f{m^{2}}{{\Lambda_{\l{MS}}}^{2}}+d)^{2}}\Big[\ldots\Big]+\ldots\bigg]\nonumber\\
&=&0\eea The total series is of course $d$-independent, but one
can only calculate it up to a certain order, this will give us a
mass gap that depends on $d$. One can check that the
$d$-dependence of the order n truncated series is
$\mathcal{O}\Big(\f{1}{\b{0}\ln
\f{m^{2}}{{\Lambda_{\l{MS}}}^{2}}+d_{0}}\Big)^{n+1}$. The only
sensible thing one can now ask for $d$ is minimal sensitivity
$\f{\partial m}{\partial d}|_{d\equiv d_{0}}=0$, and hope that the
expansion parameter $\Big(\f{1}{\b{0}\ln
\f{m^{2}}{{\Lambda_{\l{MS}}}^{2}}+d_{0}}\Big)$ is small enough to
justify the truncation.\footnote{We assume that the series is at
least asymptotic and that the size of the terms is more or less
determined by the size of the expansion parameter.}We finally note
that, as in \cite{henri},\cite{henrii}, one needs the $(n+1)$-loop
divergencies $(\gamma,\beta)$, to get an $n$-loop result.
\section{Numerical Results}\label{sec:numerical}
The exact expression for the mass gap was obtained in
\cite{forgacs}:\beb
m=(4e)^{\Delta}\f{1}{\Gamma(1-\Delta)}\Lambda_{\l{MS}},\bee where
$\Delta=\f{1}{2N-2}$. Expanding in powers of N we have:\beb
m=(1+\f{1}{N}(\f
{1-\gamma+\ln[4]}{2})+\mathcal{O}(\f{1}{N^{2}}))\Lambda_{\l{MS}}.\bee
Comparison of our results with the exact result will give us an
explicit check of our method.

The 2-loop result for $m(J)$ in the $\l{MS}$-scheme was given in
\cite{arvanitz}(section 4):\bea
m(J)&=&J\Big[1-\gg{2}\big[\f{N-1/2}{2\pi}\ln\f{J^{2}}{\mu^{2}}\big]
+\gg{4}\big[\f{(N-1/2)(N-3/4)}{4\pi^{2}}(\ln\f{J^{2}}{\mu^{2}})^{2}\nonumber\\
&&+\f{(N-1/2)(N-1/4)}{2\pi^{2}}\ln\f{J^{2}}{\mu^{2}}
+\f{N-1/2}{\pi^{2}}(0.737775-\f{\pi^{2}}{96})\big] \Big].\eea
After inversion one finds:\bea
J(m)&=&m\Big[1+\f{\gg{2}}{2\pi}\big[(N-1/2)\ln\f{m^{2}}{\mu^{2}}\big]+\f{\gg{4}}{4\pi^{2}}\big[\f{1}{4}(N-1/2)(\ln\f{m^{2}}{\mu^{2}})^{2}\nonumber\\
&&-\f{1}{2}(N-1/2)\ln\f{m^{2}}{\mu^{2}}-4(N-1/2)(0.737775-\f{\pi^{2}}{96})\big]\Big].\eea
So we have:\bea \l{\X{0}}&=&0\nonumber\\
\l{\Y{0}}&=&-\f{N-1/2}{\pi^{2}}(0.737775-\f{\pi^{2}}{96}).\eea The
$\beta$ and $\gamma$-functions have been calculated (in the
$\l{MS}$-scheme) up to three loops by Luperini and Rossi
\cite{luperini} and Gracey and Bennett \cite{gracey},
\cite{graceyii}, \cite{graceyiii}:\bea
\b{0}=\f{N-1}{2\pi},\qquad\b{1}=-\f{N-1}{4\pi^{2}},\qquad\l{\b{2}}=-\f{(N-1)(N-7/2)}{16\pi^{3}},\\
\g{0}=\f{N-1/2}{\pi},\qquad\l{\g{1}}=-\f{N-1/2}{4\pi^{2}}\qquad\l{\g{2}}=-\f{(N-1/2)(N-3/4)}{4\pi^{3}}.\eea
One can now easily check (\ref{genj}).

\subsection{one-loop results} The 1-loop result for the mass gap
is found as the solution of the 1-loop truncation of
$\widehat{J}=0$. Putting the numbers in (\ref{Jtildeln}), one
finds the mass gap as the solution of the transcendental
equation:\bea 1-\f{1}{\f{N-1}{2}\ln
\f{m^{2}}{{\Lambda_{\l{MS}}}^{2}}+d\pi}\bigg[\f{2N-1}{8(N-1)}\Big(1+4\pi
d+2\ln[\ln
\f{m^{2}}{{\Lambda_{\l{MS}}}^{2}}+\f{2d\pi}{N-1}]\Big)\bigg]=0\label{1loop}
\eea

First, the bad news: there exists no $d_{0}$ for which the
minimum-sensitivity criterium holds. Indeed, taking the derivative
to $d$ of (\ref{1loop}) gives us an equation which must also hold
if $\f{\partial m}{\partial d}=0$:\bea &&\f{\pi}{(\f{N-1}{2}\ln
\f{m^{2}}{{\Lambda_{\l{MS}}}^{2}}+d\pi)^{2}}\bigg[\f{2N-1}{8(N-1)}\Big(1+4\pi
d+2\ln[\ln
\f{m^{2}}{{\Lambda_{\l{MS}}}^{2}}+\f{2d\pi}{N-1}]\Big)\bigg]=\nonumber\\
&&\f{\pi}{\f{N-1}{2}\ln
\f{m^{2}}{{\Lambda_{\l{MS}}}^{2}}+d\pi}\bigg[\f{2N-1}{8(N-1)}\Big(4+\f{2}{\f{N-1}{2}\ln
\f{m^{2}}{{\Lambda_{\l{MS}}}^{2}}+d\pi}\Big)\bigg],\eea or using
(\ref{1loop}):\bea \f{1}{\f{N-1}{2}\ln
\f{m^{2}}{{\Lambda_{\l{MS}}}^{2}}+d\pi}=-\f{2}{(2N-1)}.\eea So we
would find a negative expansion-parameter, which is clearly not
consistent with (\ref{1loop}), since the last term gets an
imaginary value.

Now the good news: for any \textit{physical} choice for $d$ we
find reasonable results. Furthermore we can show that any physical
$d$ gives the exact $N\rightarrow\infty$ limit. This happens
because every natural coupling constant renormalization
(\ref{gtr}) leads to a value for
$\bb{0}\stackrel{N\rightarrow\infty}{=}\f{\alpha N}{\pi}$, with
$\alpha$ dependent on the condition, defining the R.S.. So if we
take $\mu=m$ in (\ref{Jtildeln}), we find
$d\stackrel{N\rightarrow\infty}{=}\f{\alpha N}{\pi}$ and equation
(\ref{1loop}) becomes:\bea
1\stackrel{N\rightarrow\infty}{=}\f{2\alpha}{\ln
\f{m^{2}}{{\Lambda_{\l{MS}}}^{2}}+2\alpha},\qquad\textrm{or
}m=\Lambda_{\l{MS}}.\eea

We now calculate the mass gap, using two different conditions for
the coupling constant renormalization. The conditions are defined
on the 1 P.I. 4-point function. In zero loops this originates from
the interactionterm $\f{\gg{2}}{2}{(\overline{\Psi}\Psi)}^{2}$. In
one loop, one finds, after a straightforward calculation, an
effective interaction of the form:
$\f{\gg{2}}{2}{(\overline{\Psi}\Psi)}^{2}\Big[1+\gg{2}f(J,s,t,u,\mu)
\Big]+\gg{4}\Big[\textrm{ \small{other operators} }\Big]$, where
$s,t,u$ are the Mandelstam-variables. Two possible natural
conditions are:\bea
f(J,s,t,u,\mu)|_{\stackrel{s=t=u=0}{\mu=J}}&=&0,\\
f(J,s,t,u,\mu)|_{\stackrel{s=t=u=-J^{2}}{\mu=J}}&=&0.\eea They
lead to two values for $d$, $d_{1}=\f{N-3/2}{\pi}$ and
$d_{2}=\f{(\f{N-1}{2}\sqrt{5}-\f{1}{\sqrt{5}})\ln\f{\sqrt{5}+1}{\sqrt{5}-1}}{\pi}$.
If we now use these values in (\ref{1loop}), we find two solutions
$m_{11}$ and $m_{12}$.

The deviation from the exact result for the $N\rightarrow\infty$
limit, the first order in $\f{1}{N}$, and our two one-loop
solutions $m_{11}$ and $m_{12}$ have been displayed in table
\ref{table1loop} in terms of percentage.
\begin{table}\label{table1loop}
\begin{center}
\begin{tabular}{|r|r|r|r|r|}
\hline N&$m_{11}$&$m_{12}$&$N=\infty$&1/N \\
\hline
2&210.7\%&268.9\%&-46.3\%&-21.9\%\\
3&31.9\%&40.6\%&-32.5\%&-12.2\%\\
4&14.7\%&19.0\%&-24.2\%&-7.0\%\\
5&9.1\%&11.9\%&-19.1\%&-4.5\%\\
6&6.5\%&8.5\%&-15.8\%&-3.1\%\\
7&5.0\%&6.6\%&-13.5\%&-2.3\%\\
8&4.0\%&5.4\%&-11.7\%&-1.8\%\\
9&3.4\%&4.5\%&-10.4\%&-1.4\%\\
10&2.9\%&3.9\%&-9.3\%&-1.1\%\\
\hline
\end{tabular}
\caption{one loop results}
\end{center}
\end{table}
Our one-loop results are acceptable for $N\geq4$. They lie
somewhere between the $N=\infty$ and $1/N$ approximations. For
$N=2,3$ the results are bad. We could try to find another
renormalization condition, which produces better results, but this
would still leave us dissatisfied. After all, the value for $d$ is
determined by some external \textit{physical} (what does that mean
anyway?) condition and does not result in a self-consistent way
from the calculations. Fortunately, this changes with the two-loop
calculations. \subsection{two-loop results} Putting the numbers in
(\ref{Jtildeln}) and equating it to zero, we find, for $N>2$, one
and only one value for $d$ (and consequently for $m$), determined
by the minimal sensitivity condition $\f{\partial m}{\partial
d}=0$. In table \ref{table2loop} we display the deviation (in
terms of percentage) from the exact result, the value for $d$ and
the value for the expansion parameter. We find excellent ($<2\%$)
agreement.

\begin{figure}
\begin{center}
\includegraphics{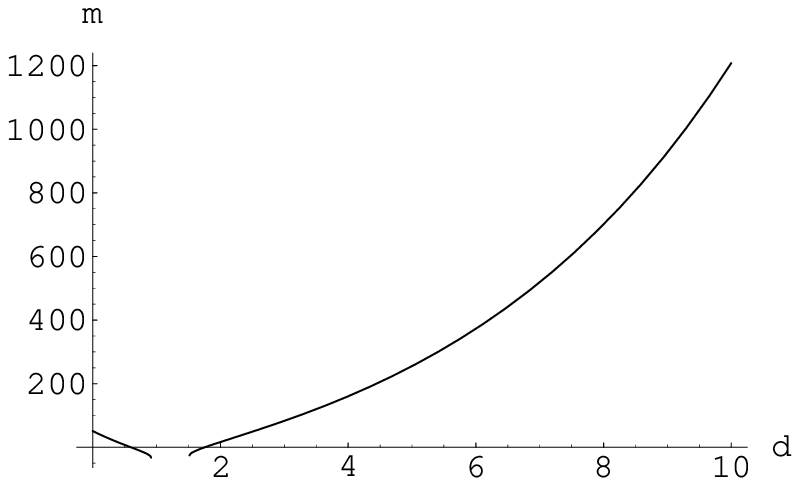}
\caption{N=2, $\pi d \rightarrow
\f{m_{2}(d)-m_{\textrm{\small{exact}}}}{m_{\textrm{\small{exact}}}}100$}
\label{plotnis2l}
\end{center}
\end{figure}

\begin{figure}
\begin{center}
\includegraphics{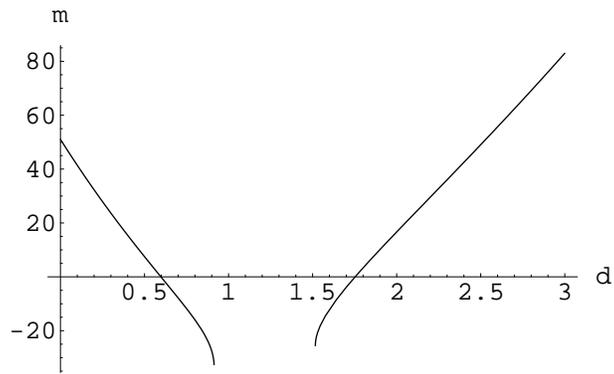}
\caption{N=2, $\pi d \rightarrow
\f{m_{2}(d)-m_{\textrm{\small{exact}}}}{m_{\textrm{\small{exact}}}}100$
, zooming in on the gap} \label{plotnis2s}
\end{center}
\end{figure}

\begin{figure}
\begin{center}
\includegraphics{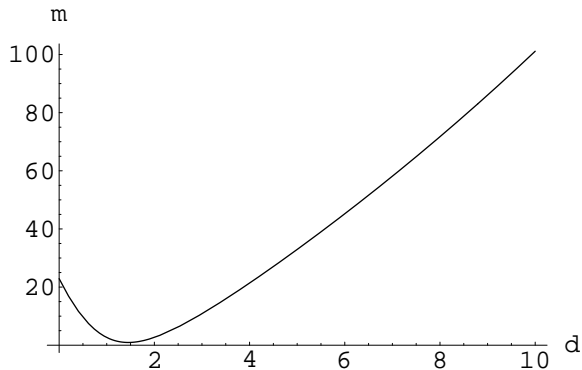}
\caption{N=3, $\pi d \rightarrow
\f{m_{2}(d)-m_{\textrm{\small{exact}}}}{m_{\textrm{\small{exact}}}}100$}
\label{plotnis3}
\end{center}
\end{figure}

The $N=2$ case is special, one finds no minimum. The graph for
$m(d)$, is displayed in figures \ref{plotnis2l} and
\ref{plotnis2s}. There is a gap with no solution for $m$ in the
region where the minimum is expected. With the minimal sensitivity
prescription in mind, we can still estimate the mass in the range
of $\pm 20\%$ deviation. The generic $N=3$ case is plotted in
figure \ref{plotnis3} for comparison.
\begin{table}\label{table2loop}
\begin{center}
\begin{tabular}{|r|r|r|c|}
\hline N&$m_{2}$&$\pi d_{0}$&$1/(\b{0}\ln [m^{2}/\Lambda_{\l{MS}}^{2}]+d_{0})\pi$\\
\hline
2&$\pm 20$\%&/&/\\
3&0.9\%&1.4&0.44\\
4&-1.0\%&2.1&0.34\\
5&-1.5\%&2.8&0.28\\
6&-1.6\%&3.4&0.24\\
7&-1.6\%&4.1&0.20\\
8&-1.5\%&4.8&0.18\\
9&-1.4\%&5.5&0.16\\
10&-1.3\%&6.1&0.14\\
15&-1.0\%&9.5&0.10\\
20&-0.8\%&12.8&0.07\\
\hline
\end{tabular}
\caption{two loop results}
\end{center}
\end{table}

\newpage
\section{Conclusions}
In this paper we have presented a general method for dynamical
mass generation in asymptotically free theories. When tested on
the Gross-Neveu model, remarkably accurate results were found for
the two-loop calculation.

Let us now try to recapture the essence of the method and its
connection with the sign of $\b{0}$. The general idea was to probe
the theory with a R.S.- and scale invariant source $\widehat{J}$.
Without the inversion one would arrive at a relation of the
following form:\bea m(\widehat{J})=\widehat{J}(\b{0}\ln
\f{{\widehat{J}}^{2}}{{\Lambda_{\l{MS}}}^{2}}+d)^{-\f{\g{0}}{2\b{0}}}\bigg[1+\f{1}{(\b{0}\ln
\f{{\widehat{J}}^{2}}{{\Lambda_{\l{MS}}}^{2}}+d)}\Big[ \ldots
\Big]\nonumber\\
+\f{1}{(\b{0}\ln
\f{{\widehat{J}}^{2}}{{\Lambda_{\l{MS}}}^{2}}+d)^{2}}\Big[\ldots\Big]+\ldots
\bigg].\label{expansion}\eea The important point is that this
expansion is only valid for large positive values of $(\b{0}\ln
m\f{\widehat{J}^{2}}{{\Lambda_{\l{MS}}}^{2}}+d)$. For an I.R.-free
theory ($\b{0}<0$) this means
$\widehat{J}^{2}\ll{\Lambda_{\l{MS}}}^{2}\exp^{-\f{d}{\b{0}}}$, so
we can take the limit $\widehat{J}\rightarrow0$ and one finds no
mass gap. For asymptotic free theories, the expansion
(\ref{expansion}) is invalid for small $\widehat{J}$, so it cannot
be used to probe the theory around $\widehat{J}=0$. The expansion
of the inverse function $\widehat J (m)$ (\ref{Jtildeln}) on the
contrary, remains well defined in the limit
$\widehat{J}\rightarrow 0$, provided that a positive solution for
m exists and that the corresponding expansionparameter $
1/(\b{0}\ln \f{{m}^{2}}{{\Lambda_{\l{MS}}}^{2}}+d)$ is not too
big. We found that this was the case for the Gross-Neveu model if
we fixed $d$ by the minimal sensitivity prescription. Calculations
on other models and on QCD are in progress.

\end{document}